\documentclass[twocolumn,fp]{jpsj3}
\usepackage{graphicx}
\usepackage{bm}
\usepackage{cite}
\usepackage{color}
\usepackage{amssymb}
\usepackage{amsmath}
\usepackage{revsymb}
\usepackage{braket}

\title{Excitonic Insulator State of the Extended Falicov--Kimball Model in the Cluster Dynamical Impurity Approximation}

\author{
Kosuke Hamada$^1$, 
Tatsuya Kaneko$^2$, 
Shohei Miyakoshi$^1$, 
and Yukinori Ohta$^1$\thanks{ohta@faculty.chiba-u.jp}
}
\inst{
$^1$Department of Physics, Chiba University, Chiba 263-8522, Japan \\
$^2$Computational Condensed Matter Physics Laboratory, RIKEN, Wako, Saitama 351-0198, Japan 
}

\abst{
We comparatively study the excitonic insulator state in the extended Falicov--Kimball model (EFKM, a 
spinless two-band model) on the two-dimensional square lattice using the variational cluster approximation 
(VCA) and the cluster dynamical impurity approximation (CDIA).  In the latter, the particle-bath sites are 
included in the reference cluster to take into account the particle-number fluctuations in the correlation 
sites.  We thus calculate the particle-number distribution, order parameter, ground-state phase diagram, 
anomalous Green's function, and pair coherence length, thereby demonstrating the usefulness of the CDIA 
in the discussion of the excitonic condensation in the EFKM.  
}

\begin{document}
\maketitle

\section{Introduction}

The excitonic phases, often referred to as excitonic insulators (EIs), are the states where the valence 
and conduction bands are hybridized spontaneously by the interband Coulomb interaction, and 
have been predicted to occur near the semimetal-semiconductor phase boundary as the quantum 
condensation of electron-hole pairs (excitons) \cite{KK65,De65,JRK67,Ko67,HR68,HR68-2}.  
In the semimetallic region, where the Coulomb interaction is largely screened by free carriers, 
the excitonic phase is described in analogy to the BCS theory of superconductivity, whereas in the 
semiconducting region, it is described as the Bose--Einstein condensation (BEC) of preformed 
excitons (or strongly bound electron-hole pairs).  Thus, the BCS-BEC crossover is expected 
to occur by controlling the band gap from a negative value to a positive one \cite{KM65,NS85,BF06}.  

In recent years, the possible realization of spin-singlet excitonic condensation has been suggested 
for transition-metal chalcogenides such as $1T$-TiSe$_2$ and Ta$_2$NiSe$_5$ 
\cite{CMCetal07,MCCetal09,vWNS10,ZFBetal13,MMHetal15,WSY15,WSTetal09,KTKetal13,SWKetal14,SKO16,SO16,YDO16,MHGetal16,LKLetal17}.  
The spin-triplet excitonic condensation has also been suggested to occur in the high-spin/low-spin 
crossover region of some cobalt oxide materials with the cubic perovskite structure 
\cite{KA14-2,INTetal16,NWNetal16,SK16,TMNetal16,YSO17}.  
Because these materials are among transition-metal chalcogenides and oxides, where the 
effects of electron correlations are strong, one must reconsider the excitonic phases 
from the standpoint of strongly correlated electron systems \cite{Ku15,KO16,OKS17}.  
Thus, the lattice models such as Hubbard models, rather than the gas models, are appropriate for use.  
The spinless extended Falicov--Kimball model (EFKM) is the simplest lattice model for describing 
the excitonic phases, and has been used to discuss, for example, the BCS-BEC crossover of the excitonic 
condensation \cite{Ba02,BGBetal04,Fa08,IPBetal08,PBF10,SEO11,ZIBetal12,KEFetal13,EKOetal14}.  
The multiband Hubbard model, taking into account the spin degrees of freedom, has also been 
used to discuss, for example, the relative stability of the spin-singlet and spin-triplet excitonic phases 
\cite{BT09-2,ZTB11,KSO12,KA14-1,Ku14,KO14,KZFetal15,KG16}. 

We have so far studied the excitonic phases of the EFKM and multiband Hubbard models 
using the variational cluster approximation (VCA) based on the exact diagonalization of small 
clusters \cite{PAD03,DAHetal04}.  However, some difficulty arises in such a small-cluster approach, 
particularly when we calculate physical quantities as a function of the model parameters.  
For example, the number of electrons in the valence and conduction bands changes discontinuously 
as a function of energy-level splitting between the valence and conduction orbitals.  This is 
because the number of electrons in the cluster is fixed, or the electron-number fluctuation in the 
cluster is prohibited, so that the finite-size effects appear even in the VCA \cite{SEO11,KSO12}, 
where the system of connecting the clusters is treated variationally in the thermodynamic limit.    

In this paper, we therefore use the cluster dynamical impurity approximation (CDIA) 
\cite{Po03-1,Po03-2,BKSetal09}, which is an extended version of the VCA, where the 
particle-bath sites are added to the reference clusters to take into account the electron-number 
fluctuations in the correlation sites.  To the best of our knowledge, the roles of the bath sites 
in the multiorbital models have not been fully elucidated, so that the present study will provide 
a useful improvement of the VCA technique for two-orbital correlated electron models.  

We will thus study the EI state of the EFKM defined on the two-dimensional square lattice, 
calculating the particle-number distribution, order parameter, ground-state phase diagram, 
anomalous Green's function, and pair coherence length.  We will show that the spurious 
discontinuities in the parameter dependence of some physical quantities, which are caused by 
the finite-size effect of the reference clusters and are inevitable in the VCA, are clearly suppressed 
in the CDIA.  We thus demonstrate the usefulness of the CDIA in the discussion of the excitonic 
condensation of the EFKM.  

This paper is organized as follows.  
In Sect.~2, we introduce the model and discuss the methods of calculation.  In Sect.~3, we show 
the calculated physical quantities characterizing the excitonic phase of the EFKM and compare 
the results of the VCA with those of the CDIA to demonstrate the usefulness of the CDIA.  
A summary of the paper is given in Sect.~4.  

\section{Model and Method}

\subsection{Extended Falicov--Kimball Model}

In this paper, we use the EFKM, which is the simplest lattice model to discuss the EI state 
\cite{Ba02,BGBetal04,Fa08,IPBetal08,PBF10,SEO11,ZIBetal12,KEFetal13,EKOetal14}.  
Assuming a two-dimensional square lattice, we define the Hamiltonian of the EFKM as 
\begin{align}
\mathcal{H}
=-t_{f}&\sum_{\langle i,j\rangle}f^{\dag}_{i}f_{j}^{}-t_{c} \sum_{\langle i,j\rangle}c^{\dag}_{i} c_{j} 
+\frac{D}{2} \sum_{i} \left( n_{ic} - n_{if} \right)
\notag \\
 &+U\sum_{i}n_{if}n_{ic} - \mu \sum_{i} \left( n_{if} + n_{ic} \right) , 
\label{efkm-cdia-eq1}
\end{align} 
where $f^{\dag}_i$ ($c^{\dag}_i$) and $f_i$ ($c_i$) denote the creation and annihilation operators 
of an electron on the $f$ ($c$) orbital at site $i$, respectively, and $n_{if} = f^{\dag}_{i}f_{i}$ 
($n_{ic} = c^{\dag}_{i}c_{i}$).  $t_f$ ($t_c$) is the transfer integral between the $f$ ($c$) orbital 
on the neighboring sites and $D$ is the level splitting.  Here, we assume $D \ge 0$, so that the 
$f$ and $c$ bands correspond to the valence and conduction bands, respectively.  
$U$ $(>0)$ is the interorbital Coulomb repulsion, leading to the excitonic instability in the system.  
The chemical potential $\mu$ is determined to maintain the total particle density at half-filling: 
$\langle n_f \rangle + \langle n_c \rangle = 1$.  

It is known that the EFKM at half-filling contains three phases: the staggered orbital ordered (SOO) 
phase, the EI phase, and the band insulator (BI) phase \cite{Ba02,BGBetal04,Fa08,EKOetal14}.  
At $D=0$, where $\langle n_f \rangle = \langle n_c \rangle = 0.5$, the SOO phase appears, 
in which an electron occupies the $c$ and $f$ orbitals alternately.  Comparison with the XXZ model 
obtained in the strong correlation limit of the EFKM indicates that the SOO phase becomes more 
stable when $|t_c|/|t_f| \ll 1$ \cite{BGBetal04,EKOetal14}.  On the other hand, when $D \gg |t_f|, |t_c|$, 
the BI phase appears, where $\langle n_f \rangle =1$ and $\langle n_c \rangle =0$.  
Between these two phases, the EI phase appears, where the $f$ and $c$ orbitals hybridize 
spontaneously owing to the interaction $U$.  With increasing $U$ from the noninteracting semimetallic 
state, the EI state shows a smooth crossover from the weak-coupling BCS state to the strong-coupling 
BEC state \cite{IPBetal08,PBF10,SEO11,ZIBetal12}.  Thus, the EFKM is used to discuss the 
BCS-BEC crossover of the excitonic condensation.  

Hereafter, we assume $t_f = t_c = t$ and use $t$ as the unit of energy.  
When $t_f=t_c$, the SOO and EI phases have the same energy at $D=0$, but at $D>0$, the degeneracy 
is lifted and the EI phase becomes the most stable one \cite{BGBetal04,EKOetal14}.  
Note that the conduction-band bottom is located at $\bm{k} = (0, 0)$, giving rise to an electron 
pocket, while the valence-band top is located at $\bm{k} = (\pi, \pi)$, giving rise to a hole pocket.  
Thus, the EI state of the present model is characterized by the modulation vector $\bm{Q}=(\pi, \pi)$.  
Throughout the paper, we set $\hbar = 1$ and lattice constant $a = 1$.
We fix the chemical potential at $\mu=U/2$ because of the particle-hole symmetry at $t_c=t_f$.  

\subsection{VCA and CDIA}

To accomplish the calculations treating electronic correlations accurately in the thermodynamic limit, 
we employ the VCA and CDIA, which are quantum cluster methods based on the self-energy functional 
theory (SFT) \cite{Po03-1,Po03-2,PAD03,DAHetal04}. 
In the SFT, the grand potential $\Omega$ of the original system is given by a functional of the self-energy.  
In the VCA and CDIA, we introduce disconnected finite-size clusters that are solved exactly as a reference 
system.  By restricting the trial self-energy to that of the reference system $\hat{\Sigma}^{\prime}$, 
we obtain the grand potential in the thermodynamic limit as 
\begin{align}
\Omega[\hat{\Sigma}']
&=\Omega'+\mathrm{Trln}(\hat{G}_0^{-1}-\hat{\Sigma}')^{-1}-\mathrm{Trln}(\hat{G}') 
\notag \\
&=\Omega' - \mathrm{Trln}(\hat{I}-\hat{V}\hat{G}') , 
\label{efkm-cdia-eq2}
\end{align}
where $\Omega^{\prime}$ and $\hat{G}'$ [$=({\hat{G}^{\prime-1}_0} -\hat{\Sigma}')^{-1}$] are the exact 
grand potential and Green's function of the reference system, respectively, and $\hat{G}_0$ ($\hat{G}'_0$) 
is the noninteracting Green's function of the original (reference) system.  $\hat{I}$ is the unit matrix and 
$\hat{V} = \hat{G}^{\prime-1}_0-{\hat{G}_0}^{-1}$ corresponds to the hopping parameter between 
the adjacent clusters.  In this method, the short-range electron correlations within the cluster of the 
reference system are taken into account exactly.  
Details of the method can be found in Refs.~55 and 56.  

In our calculations, when a Hamiltonian of a reference system is given as $\mathcal{H}'$, we solve the 
eigenvalue problem $\mathcal{H}'|\psi_0\rangle = E_0|\psi_0\rangle$ of a finite-size ($L_c$ sites) cluster 
to obtain the ground state, and thus, the trial Green's function is calculated by the Lanczos 
exact-diagonalization method.  The EFKM is a two-orbital model, and using the basis 
$(f^{\dag}_{i},c^{\dag}_{i})$, the Green's function matrix $\hat{G}'$ in Eq.~(\ref{efkm-cdia-eq2}) may be 
written as 
\begin{align}
\hat{G}'(\omega )
=\left( \begin{array}{cc}
\hat{G}'^{ff}(\omega ) &\hat{G}'^{fc}(\omega ) \\
\hat{G}'^{cf}(\omega ) &\hat{G}'^{cc}(\omega )  \\
\end{array} \right) , 
\end{align}
where $\hat{G}'^{\alpha\beta}$ is an $L_c \times L_c$ matrix, and each matrix element is defined as 
\begin{align}
G'^{\alpha\beta}_{ij}(\omega) 
&= \langle \psi_0 | \alpha_{i}\frac{1}{\omega-\mathcal{H}'+E_0}\beta^{\dag}_{j} |\psi_0\rangle  \nonumber\\
&+ \langle \psi_0 | \beta^{\dag}_{j} \frac{1}{\omega+\mathcal{H}'-E_0}\alpha_{i} |\psi_0\rangle.
\end{align} 

\begin{figure}[t]
\begin{center}
\includegraphics[width=\columnwidth]{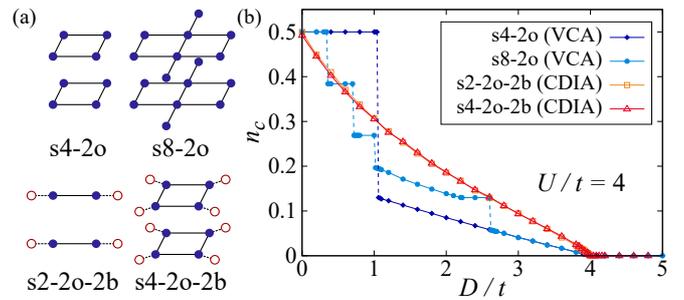}
\caption{(Color online) 
(a) Schematic representations of the reference clusters used in this paper.  
The 4-site 8-orbital (s4-2o), 
8-site 16-orbital (s8-2o), 
2-site 4-orbital 4-bath (s2-2o-2b), and 
4-site 8-orbital 8-bath (s4-2o-2b) clusters are illustrated.   
(b) Calculated number of particles $n_c$ as a function of $D$ at $U/t=4$ obtained by the VCA and CDIA, 
where the normal phase without the EI order is assumed.
}\label{fig1}
\end{center}
\end{figure}

The advantage of the VCA is that the spontaneous symmetry breaking can be treated within the framework 
of the theory \cite{DAHetal04}.  In this paper, we consider the EI state of the EFKM, so that the Hamiltonian 
of the variational Weiss field is given as \cite{SEO11} 
\begin{align}
\mathcal{H}'_{\mathrm{EI}}
&= \Delta' \sum_{\bm{k}} c^{\dag}_{\bm{k}+\bm{Q}}f_{\bm{k}}+\mathrm{H.c.}
\notag \\
&= \Delta' \sum_{i}  e^{i\bm{Q}\cdot\bm{r}_i}\ c^{\dag}_{i}f_{i}+\mathrm{H.c.} , 
\end{align} 
where $\bm{Q} = (\pi,\pi)$ in our indirect gap system ($t_f=t_c$).
Thus, in our VCA calculations, the Hamiltonian of the reference system is given as 
$\mathcal{H}' = \mathcal{H} + \mathcal{H}'_{\mathrm{EI}}$ 
and the variational parameter $\Delta'$ is optimized on the basis of the variational principle,  
$\partial\Omega/\partial\Delta'=0$.  
The solution with $\Delta' \ne 0$ corresponds to the EI state.  
We use the 4-site 8-orbital cluster (s4-2o) and 8-site 16-orbital cluster (s8-2o) shown in Fig.~\ref{fig1}(a) 
as the reference systems, where 2o denotes 2 orbitals per site.  

A problem with the VCA is that the effects of the particle-number fluctuations are not fully taken into 
account because the number of electrons in the cluster of the reference systems is fixed.  
In particular, in the EFKM, the numbers of electrons in the $f$ and $c$ orbitals, $N_f$ and $N_c$, respectively, 
which satisfy $N_f + N_c =L$, change discontinuously as a function of $D$ when we calculate the ground 
state using the reference system of the exactly solved finite-size clusters.  
Although this discontinuity is partially relaxed when we introduce the variational parameter of the EI that 
yields the hybridization between the $f$ and $c$ orbitals, the effects of using the finite-size clusters in 
the reference system inevitably affect the results of the VCA calculations.  Thus, the VCA, although formulated 
for calculations in the thermodynamic limit, cannot avoid the finite-size effects of the reference clusters. 

To solve this problem, we use the CDIA \cite{BKSetal09,HS13,EKO15}, which is an extended version of the 
VCA, where the particle-bath sites are added to the clusters to take into account the electron-number 
fluctuations in the correlation sites, and is intrinsically equivalent to the cluster dynamical mean-field 
theory with an exact-diagonalization solver \cite{Se12-2}.  
In the CDIA, we introduce the hybridization parameter between the bath and correlation sites. 
We use the 2-site 4-orbital correlation cluster with 4 bath sites (s2-2o-2b) and 4-site 8-orbital correlation 
cluster with 8 bath sites (s4-2o-2b) as the reference systems, where 2o denotes 2 orbitals per site 
and 2b denotes 2 bath sites per correlation site, as shown in Fig.~\ref{fig1}(a).  
In these reference systems, the variational Hamiltonian for the hybridization between the bath and 
correlation sites is given as 
$\mathcal{H}'_{\mathrm{hyb}}= V \sum_{i}  \big( a^{\dag}_{if} f_i + a^{\dag}_{ic} c_i  + \mathrm{H.c.} \big)$, 
where $a^{\dag}_{if}$ ($a^{\dag}_{ic}$) is the creation operator of an electron at bath site $i$ for the 
$f$ ($c$) orbital.  The hybridization parameter $V$ is optimized on the basis of SFT \cite{BKSetal09}.  
In the CDIA, we must introduce the Weiss field for the EI in the bath sites, as well as in the 
correlation sites \cite{Se12-2}.  The Weiss field in the present case is given by 
$\mathcal{H}'_{\mathrm{EI,b}}= - \Delta' \sum_{i}  e^{i\bm{Q}\cdot\bm{r}_i}\ a^{\dag}_{ic}a_{if}+\mathrm{H.c.}$, 
which enables us to obtain physically meaningful solutions.  
Thus, in our CDIA calculations, the Hamiltonian of the reference system is given by 
$\mathcal{H}' = \mathcal{H} +\mathcal{H}'_{\mathrm{hyb}} + \mathcal{H}'_{\mathrm{EI}} + \mathcal{H}'_{\mathrm{EI,b}}$, 
whereby we optimize the parameters $V$ and $\Delta'$ by applying the variational principle.  

\section{Results of Calculations}


First, let us discuss the number of particles in the normal state ($\Delta'=0$) obtained using the VCA and CDIA.  
The average number of particles on the $\alpha$ (=$f$, $c$) orbital is given as 
\begin{align}
n_{\alpha}=\frac{1}{L}\sum_{i} \langle \alpha^{\dag}_i \alpha_i \rangle
=\frac{1}{N_cL_c}\oint_{C}\frac{{\rm d}z}{2\pi i}\sum_{\bm{K}}\sum^{L_c}_{l=1}\mathcal{G}^{\alpha\alpha}_{ll}(\bm{K},z) , 
\end{align}
where $\hat{\mathcal{G}}^{\alpha\alpha}(\bm{K},z)$ is the diagonal block (of size $L_c \times L_c$) of the full 
matrix $\hat{\mathcal{G}}(\bm{K},z)=\big[\hat{G}'^{-1}(z)-\hat{V}(\bm{K})\big]^{-1}$ (of size $2L_c \times 2L_c$).  
The $\bm{K}$-summation is performed in the reduced Brillouin zone of the superlattice, $N_c$ is the number of 
clusters, and the contour $C$ of the frequency integral encloses the negative real axis.  
In Fig.~\ref{fig1}(b), we show the calculated results for the number of particles in the conduction band $n_c$ 
as a function of $D$.  We find that, in both the VCA and CDIA, the value of $n_c$ decreases from 0.5 at $D=0$ to 0 
at $D\simeq 4$, resulting in the BI at $D\agt 4$.  We then find that, in the VCA, the value of $n_c$ changes discontinuously as a function 
of $D$, reflecting the discontinuous change in the number of particles obtained by the small-cluster diagonalization.  
However, in the CDIA, where the bath sites are introduced and the particle-number fluctuations are allowed, we 
find that the value of $n_c$, optimized by the variational calculation, decreases smoothly as a function of $D$, 
as is expected in the thermodynamic limit.  We also find that the results for $n_c$ depend little on the choice 
of the reference clusters s2-2o-2b and s4-2o-2b.  The CDIA thus works very well for the present model.  

\begin{figure}[tb]
\begin{center}
\includegraphics[width=\columnwidth]{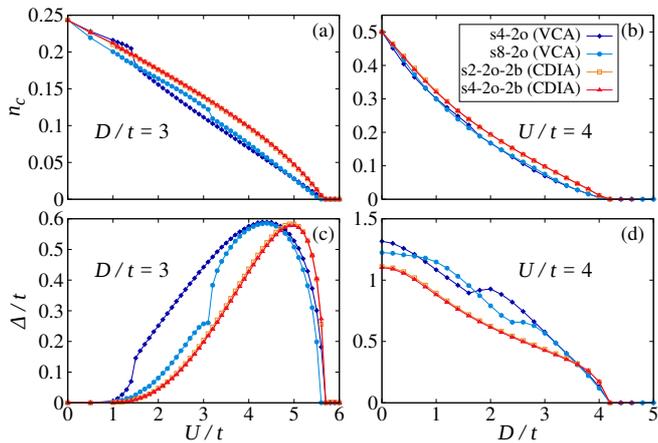}
\caption{(Color online)
Calculated results for the particle number $n_c$ and the order parameter $\Delta$ in the EI state.  
Results of the VCA and CDIA are shown for comparison.  $U$ dependences are shown at $D/t=3$ in 
(a) and (c) and $D$ dependences are shown at $U/t=4$ in (b) and (d).  
}\label{fig2}
\end{center}
\end{figure}

Next, let us discuss the number of particles and the order parameter in the EI state ($\Delta'_{\rm opt}\ne 0$).  
The order parameter of the EI state is defined as 
\begin{align}
\Delta
&=\frac{U}{L}\sum_{\bm{k}} \langle c^{\dag}_{\bm{k}+\bm{Q}} f_{\bm{k}} \rangle
=\frac{U}{L}\sum_{i} \langle c^{\dag}_i f_i \rangle e^{i\bm{Q}\cdot\bm{r}_i}
\notag \\
&=\frac{U}{N_cL_c}\oint_{C}\frac{{\rm d}z}{2\pi i}\sum_{\bm{K}}\sum^{L_c}_{l=1}\mathcal{G}^{cf}_{ll}(\bm{K},z) e^{i\bm{Q}\cdot\bm{r}_l} , 
\end{align}
where $\hat{\mathcal{G}}^{cf}(\bm{K},z)$ is the off-diagonal block (of size $L_c \times L_c$) 
of the full matrix $\hat{\mathcal{G}}(\bm{K},z)$.  

The number of particles $n_c$ and the order parameter $\Delta$ calculated by the VCA and CDIA are 
shown in the Fig.~\ref{fig2}.  
Here, we first refer to some general aspects of the EI phase of the EFKM.  
We note in Fig.~\ref{fig2}(a) that with increasing $U$, the effective energy-level difference increases 
owing to the Hartree shift, resulting in the BI phase at a large $U$.  
We also note in Fig.~\ref{fig2}(c) that the order parameter increases with increasing $U$, reaches a peak 
at a certain $U$ value, and then decreases rapidly as the system approaches the BI phase.  
This is a manifestation of the crossover between the BCS-like weak-coupling EI phase at small $U$ 
and the BEC-like strong-coupling EI phase at large $U$ \cite{IPBetal08,PBF10,SEO11,ZIBetal12}.  
We, moreover, note in Fig.~\ref{fig2}(d) that the order parameter increases with decreasing $D$ because 
the hole Fermi surface of the valence band and electron Fermi surface of the conduction band, which 
contribute to the formation of the EI phase, both become large with decreasing $D$.  
This is particularly the case when the Fermi surface nesting at the wave number $\bm{Q}$ is perfect 
and the screening of the Coulomb interaction is absent, as in the present model.  Imperfect Fermi surface 
nesting would change the present result considerably.  

\begin{figure}[bth]
\begin{center}
\includegraphics[width=0.9\columnwidth]{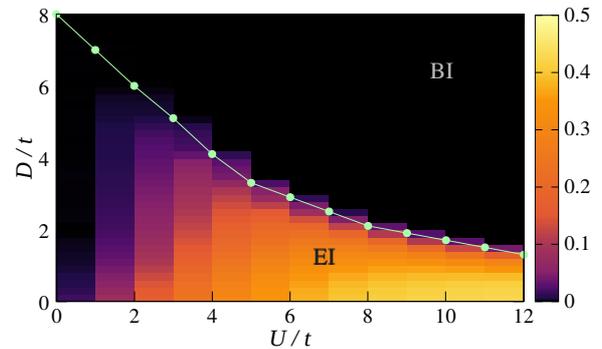}
\caption{(Color online)
Calculated ground-state phase diagram of the EFKM in the CDIA.  
The order parameter $\Phi=\Delta/U$ is indicated by color.  
The reference cluster s4-2o-2b was used.  
}\label{fig3}
\end{center}
\end{figure}

Now, let us compare the results of the VCA with those of the CDIA.  
We first find in Figs.~\ref{fig2}(a) and \ref{fig2}(b) that the discontinuous changes in $n_c$ calculated 
in the VCA are weaker than those obtained for the normal state discussed above [see Fig.~\ref{fig1}(b)].  
This is because the nonvanishing variational parameter of the EI state, which corresponds to the band 
hybridization, allows the particle-number fluctuations between the $f$ and $c$ orbitals.  
However, we still find the discontinuities in $n_c$ calculated in the VCA [see Fig.~\ref{fig2}(a)], which 
are due to the finite-size effect of the reference cluster.  
The discontinuities are more serious in the calculated order parameter in the VCA, as shown in Fig.~\ref{fig2}(c).  
The order parameter changes rather smoothly in the boundary region near the BI phase, where the 
band overlap is small, but in the region where the band overlap is large and the discontinuities in 
$n_c$ are strong, the finite-size effect appearing in the order parameter becomes serious.  
In the CDIA, on the other hand, we find that the finite-size effects in the VCA are suppressed 
completely owing to the particle-number optimization by the bath sites, so that, using the CDIA, we 
can calculate the order parameter and $n_c$ that change smoothly in a wide parameter space.  
Using the CDIA, we can thus calculate the order parameter in the EI state by suppressing the 
finite-size effects clearly.  The ground-state phase diagram and order parameter $\Phi=\Delta/U$ 
in the CDIA are then calculated and the results are shown in the $U$--$D$ plane in Fig.~\ref{fig3}.  
We find that the transition point $D_c$ between EI and BI phases decreases with increasing 
$U$ and that the order parameter becomes large in the large-$U$ region below $D_c$.  

\begin{figure}[thb]
\begin{center}
\includegraphics[width=0.8\columnwidth]{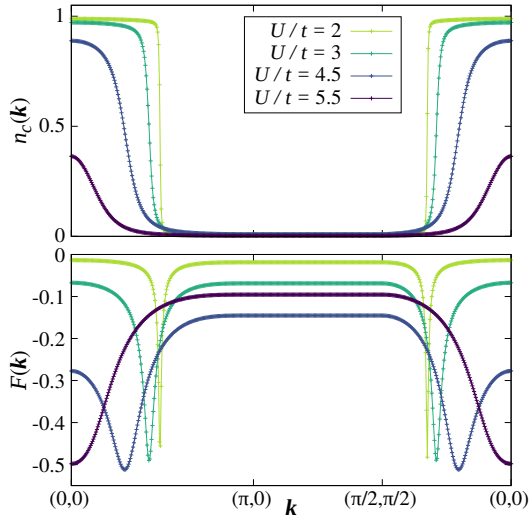}
\caption{(Color online)
Calculated results for the electron momentum distribution function $n_c(\bm{k})$ and 
anomalous momentum distribution function $F(\bm{k})$ in the CDIA.  
The reference cluster s4-2o-2b was used and $D/t=3$ was assumed.  
}\label{fig4}
\end{center}
\end{figure}


Finally, to discuss the BCS-BEC crossover of the EI state in the EFKM, we calculate the particle momentum 
distribution function $n_{\alpha}(\bm{k})$ and excitonic anomalous momentum distribution function 
$F(\bm{k})$ \cite{SEO11,KSO12} using Green's function in the cluster perturbation theory (CPT) \cite{SPP00}, 
as given by 
\begin{align}
\hat{\mathcal{G}}_{\rm CPT}(\bm{k},\bm{k}',\omega) = 
\frac{1}{L_c}\sum_{i,j} \hat{\mathcal{G}}_{ij}(\bm{k},\omega)e^{-i\bm{k}\cdot\bm{r}_i + i\bm{k}'\cdot\bm{r}_j} , 
\end{align}
where $\hat{\mathcal{G}}_{\rm CPT}$ is a $2 \times 2$ matrix.  
From this Green's function, the particle and anomalous momentum distribution functions are calculated as 
\begin{align}
n_{\alpha}(\bm{k})
&=\oint_{C}\frac{{\rm d}z}{2\pi i} \mathcal{G}^{\alpha\alpha}_{\rm CPT}(\bm{k},\bm{k},z) 
\\
F(\bm{k})
&=\oint_{C}\frac{{\rm d}z}{2\pi i} \mathcal{G}^{cf}_{\rm CPT}(\bm{k},\bm{k}+\bm{Q},z), 
\end{align}
respectively.  
Using $F(\bm{k})$, we may also evaluate the pair coherence length $\xi$, the spatial size of the electron-hole pair, 
as \cite{SEO11,ONEetal95,KO14JPSJ} 
\begin{align}
\xi^2
=\frac{\sum_{\bm{k}}|\nabla_{\bm{k}}F(\bm{k})|^2}{\sum_{\bm{k}}|F(\bm{k})|^2}.  
\end{align}

\begin{figure}[htb]
\begin{center}
\includegraphics[width=0.8\columnwidth]{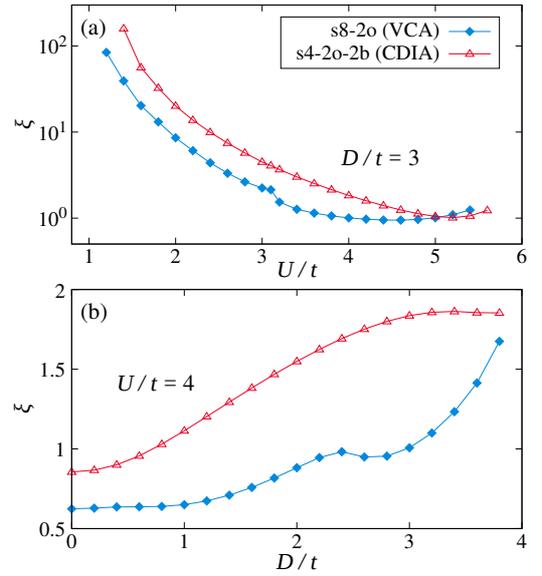}
\caption{(Color online)
Pair coherence length $\xi$ (in units of the lattice constant) calculated in the VCA and CDIA.  
(a) $U$ dependence at $D/t=3$ and (b) $D$ dependence at $U/t=4$ are shown.  
}\label{fig5}
\end{center}
\end{figure}  

The calculated results for the momentum distribution functions $n_{c}(\bm{k})$ and $F(\bm{k})$ are 
shown in Fig.~\ref{fig4}, and those of the pair coherence length $\xi$ are shown in Fig.~\ref{fig5}.  
We find in Fig.~\ref{fig4} that for small $U$, $n_c(\bm{k})$ changes rapidly at the Fermi momentum 
$\bm{k}_{\rm F}$ defined by $n_c(\bm{k}_{\rm F})=0.5$, but for large $U$ near the BI phase, 
$n_c(\bm{k})$ becomes small and broad in momentum space.  
We also find in Fig.~\ref{fig4} that $F(\bm{k})$ shows a sharp peak at $\bm{k}_{\rm F}$ for small $U$, 
which broadens with increasing $U$.  The sharp (broad) peak in $F(\bm{k})$ indicates that 
the size of the electron-hole pair is large (small) in real space \cite{SEO11}.  
This is evident in the calculated pair coherence length $\xi$ shown in Fig.~\ref{fig5}; i.e., $\xi$ is 
much larger (smaller) than the lattice constant for small (large) $U$.  Thus, the BCS-BEC 
crossover occurs between the weak-coupling small-$U$ region and the strong-coupling large-$U$ region.  
Also noted in Fig.~\ref{fig5} is that $\xi$ shows a discontinuous or kinklike structure in the VCA, 
which is clearly suppressed in the CDIA, similarly to the results for the particle number and order 
parameter shown in Fig.~\ref{fig2}.  These results thus demonstrate the usefulness of the CDIA in 
discussing the excitonic condensation in the EFKM.  

\section{Summary}

We studied the EI state of the EFKM defined on the two-dimensional square lattice, and calculated the 
particle-number distribution, order parameter, ground-state phase diagram, anomalous Green's function, 
and pair coherence length.  We used the VCA and CDIA comparatively.  We showed that the 
spurious discontinuities appearing in the parameter dependence of some physical quantities in the 
VCA are clearly suppressed in the CDIA, thus demonstrating that the addition of the particle-bath 
sites to take into account the electron-number fluctuations in the correlation sites provides an 
essential improvement of the VCA technique when we discuss the excitonic condensation of the EFKM.  

We hope that the present study on the roles of the bath sites will offer useful improvements of the 
quantum cluster methods, such as the VCA, when they are applied to the multiorbital models.  

\medskip
\begin{acknowledgment}
We thank K.~Seki and K.~Sugimoto for enlightening discussions.  This work was supported in part by 
a Grant-in-Aid for Scientific Research (No. 26400349) from JSPS of Japan.  T.~K. and S.~M. acknowledge 
support from the JSPS Research Fellowship for Young Scientists.  
\end{acknowledgment}

\end{document}